# Securing SMS Based One Time Password Technique from Man in the Middle Attack

Safa Hamdare, Varsha Nagpurkar, Jayashri Mittal
*Assistant Professor, Department of Computers, SFIT*
*Mumbai, India.*

*Abstract*- Security of financial transactions in E-Commerce is difficult to implement and there is a risk that user's confidential data over the internet may be accessed by hackers. Unfortunately, interacting with an online service such as a banking web application often requires certain degree of technical sophistication that not all Internet users possess. For the last couple of year such naive users have been increasingly targeted by phishing attacks that are launched by miscreants who are aiming to make an easy profit by means of illegal financial transactions. In this paper, we have proposed an idea for securing e-commerce transaction from phishing attack. An approach already exists where phishing attack is prevented using one time password which is sent on user's registered mobile via SMS for authentication. But this method can be counter attacked by "Man in the Middle". In our paper, a new idea is proposed which is more secure compared to the existing online payment system using OTP. In this mechanism OTP is combined with the secure key and is then passed through RSA algorithm to generate the Transaction password. A Copy of this password is maintained at the server side and is being generated at the user side using a mobile application; so that it is not transferred over the insecure network leading to a fraudulent transaction.

## I. INTRODUCTION

E-commerce is a term for any type of business, or commercial transaction that provides services for buying and selling products or exchanges information across the internet. The various E-commerce applications are online shopping, banking, e-tickets etc. An online transaction system is a payment method that authorizes transfer of funds over an Electronic Fund Transfer (EFT). In online transactions, consumers use their credit or debit card to buy products from seller [3]. In most of the countries, people are using these services for sale and purchase. Credit and Debit cards have increasingly become the preferred mode of payments for goods and services, making these forms of electronic payments an indispensable way for merchants big and small to conduct business [2]. Online services simplify our lives. They allow us to access information ubiquitously and are also useful for service providers because they reduce the operational costs involved in offering a service. For example, online banking over the web has become essential for customers as well as for banks. In spite of all these benefits, e-commerce is not completely secure and has few inherent risks. Let us look at one of the major risks in e-commerce which has been addressed in this paper is "Phishing".

## II. ANALYSIS OF PHISHING

Phishing is a form of electronic identity theft in which a combination of social engineering and web site spoofing techniques are used to trick a user into revealing confidential information with economic value. In a typical attack, the attacker sends a large number of spoofed e-mails to random internet users that appear to be coming from a legitimate business organization such as a bank. The e-mail urges the recipient (i.e., the potential victim) to update his personal information using links in the email, if the recipient does not do so it will result in the suspending of his online banking account.

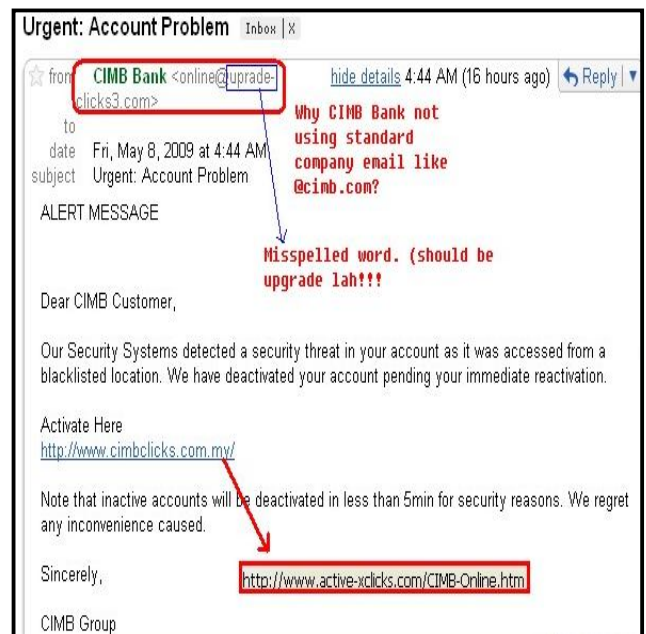
Fig 1: Spoofed email for Phishing attack

Such un-grounded threats are common in social engineering attacks and are an effective technique in persuading users. When the unsuspecting victim follows the phishing link provided in the e-mail, he is directed to a web site that is under the control of the attacker. The site is prepared in a way such that it looks familiar to the victim by imitating the visual corporate identity of the target organization by using same icons, logos and textual Info.

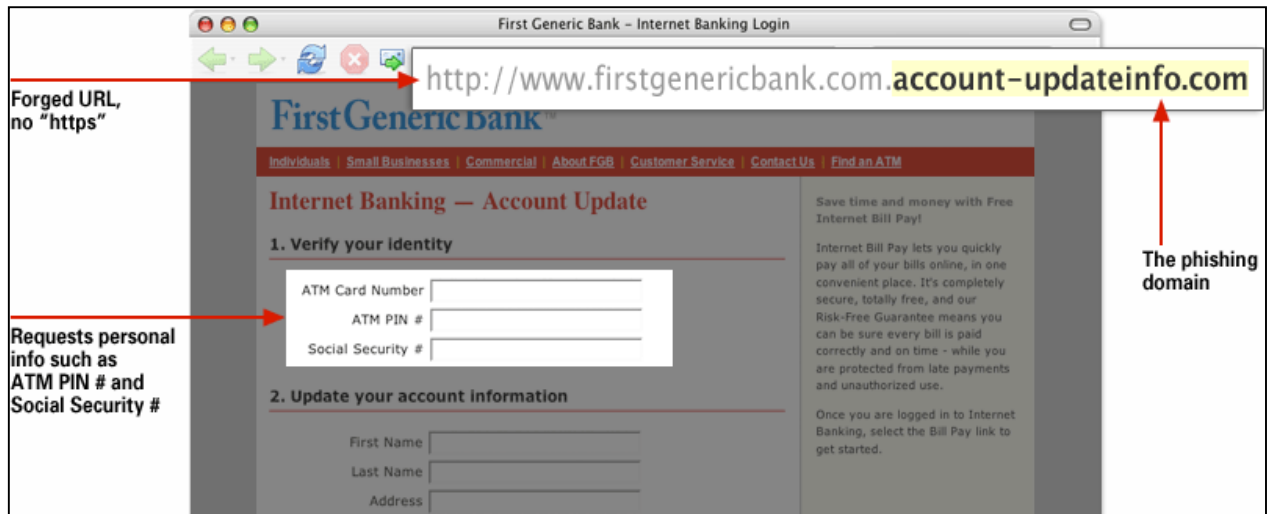

Fig 2: An Example showing Phishing attack by visual Deception requesting Personal Bank Information

**Phishing by Visual Deception [6]:**
1. **Visually deceptive text:** Users may be fooled by the syntax of a domain name in "type jacking" attacks, which substitute letters that may go unnoticed (e.g. www.paypai.com uses a lowercase "i" which looks similar to the letter "l", and www.paypa1.com substitutes the number "1" for the letter "l".

2. **Images masking underlying text:** One common technique used by phishers is to use an image of a legitimate hyperlink. The image itself serves as a hyperlink to a different, rogue site.

3. **Windows masking underlying windows:** A common phishing technique is to place an illegitimate browser window on top of, or next to, a legitimate window. If they have the same look and feel, users may mistakenly believe that both windows are from the same source, regardless of variations in address or security indicators. In the worst case, a user may not even notice that a second window exists (browsers that allow borderless pop-up windows aggravate the problem).

4. **Deceptive look and feel:** If images and logos are copied perfectly, sometimes the only cues that are available to the user are the tone of the language, misspellings or other signs of unprofessional design. If the phishing site closely mimics the target site, the only cue to the user might be the type and quantity of requested personal information.

### III. PHISHING PREVENTION TECHNIQUE

Many solutions have been developed to fight against phishing attacks. Yet, the tendency of phishing is expanding and a number of new techniques are being implemented to cause more damage, it has thus become a serious cybercriminal activity. Most of the cybercrimes are caused because of traditional, static password which is usually only changed when necessary: either when it has expired or when the user has forgotten it and needs to reset it. As passwords are cached on computer hard drives and stored on servers, they are susceptible to "cracking". This is especially a concern for people accessing their bank accounts from different locations and sometimes through an unsafe network. Unlike a static password, one-time password changes each time the user logs-in and thus can be a solution for the above mentioned problem.

One Time passwords as the name suggests are passwords that can be used only one time, you need not remember the same. It has an expiry Time on how long after it has been issued, the password remains valid. It can be valid for 5 minutes only. One time passwords are not vulnerable to replay attacks, since the password cannot be used again. It is sent to a person's registered cell phone via SMS, Hence the name SMS based OTP and they are expected to enter it on a website. Short Message Service (SMS) based One-Time Passwords (OTP) were introduced to counter phishing attacks against authentication and authorization of Internet services [1].

There are few reasons why this is a very strong method. These few characteristics make the OTP a strong authentication protocol.

- A password is valid once and for a very short period of time
- The algorithm that generates each password is not reversible
- With an OTP token, the client Status is verified
- As the OTP is received on client's phone, the user is authenticated again

A. **One Time Password Generation Technique [1]:**

   This Section presents the implementation of OTP Generation and how it can help in mitigating a typical phishing attempt. Whenever user wishes to do net banking, First step is to enter user ID and Pin number for User authentication. Once user is authenticated he initiates his transaction and gets the One Time password by SMS on his registered mobile number and a token is generated which is automatically stored on user's machine. Token with status value 1 is valid which signifies that OTP can still be used by the user. The moment user uses the generated OTP or after a period of 5 minutes since user received the OTP on mobile, the OTP expires and its token value changes from 1 to 0. There is one more reason for generating token with OTP i.e. as it is stored on user machine it will authenticate that OTP has been sent through the primary web site by the same client who has initiated the transaction.

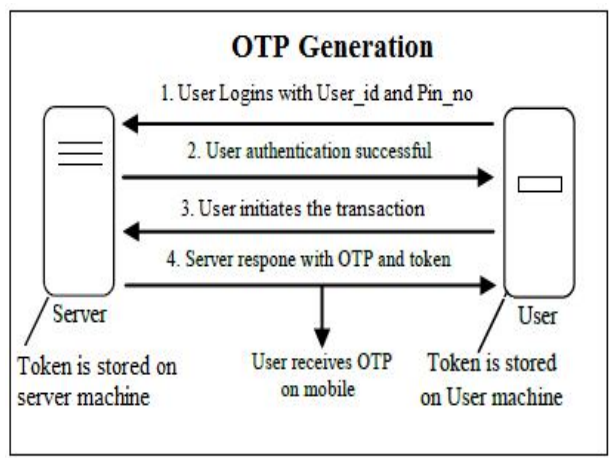

Fig. 3: Getting one time password and authenticating user

   When the user enters the OTP which he has received on mobile it is checked with server side OTP along with that, token value and its status is also compared for successful authentication. The cookie is valid only for 5 minutes and contains user machine IP address and other details. However, if attacker is able to get the user credentials by forged website (through Phishing attack), he/she will not be able to cause any damage as the transaction is not complete without OTP which is only accessible to the valid user as it is sent on the registered mobile phone.

After the Generation of OTP the details that are preserved with server is as followed:

TABLE I: Login table to the main website with OTP

| Username | One Time Password | Token | Status |
|---|---|---|---|
| John | 891632 | 1 | Valid |
| Henry | 764589 | 0 | Expired |
| Russell | 123141 | 1 | Valid |
| David | 456723 | 0 | Expired |

B. **Vulnerabilities of SMS OTP Authentication:**

   Currently, almost all transactions made via Master/Visa credit cards or banking transactions require an additional One Time Password (OTP) SMS Verification. This requirement has been made mandatory by banks for quite some time. The introduction of OTP is intended to reduce the possibility of fraudulent transactions and safeguard customers. However, SMS OTP as a form of online security is fraught with few deficiencies and is not difficult to hack.

   SMS OTP is in-band authentication which relies on a single mode of communication between a user and the relying web service. Because of the single stream, fraudsters can attack in-band authentication with Man in the Middle (MITM) attacks (phishing) [8]. **Man in the middle (MITM)** attack infects the mobile device and sniffs all the SMS messages that are being delivered to user to a server controlled by attacker. Suppose through phishing attack, attacker has stolen both the online user ID and Pin Number. Now the only security that is currently blocking the phishing attack is the SMS OTP which user receives on mobile device.

   As part of MITM attack, the attacker infects the user's mobile device by forcing him to install a malicious application (he sends a SMS with a link to the malicious mobile application). If user clicks on the link, malicious software gets installed on user's phone which forwards the SMS on user's phone to other terminal/server controlled by the attacker. If the attacker logs in with the stolen credentials i.e. userid and pin number which he already had, this operation sends an OTP on customers mobile through an SMS. The malicious software forwards the SMS to terminal/server controlled by the attacker. The attacker having customer's credentials now also has access to the OTP and can carry out a fraudulent transaction. In such a scenario, even the token value cannot stop this attack as the transaction was initiated by the malicious user on website and server has stored the token value in the attacker's machine.

IV. **METHODOLOGY FOR SECURING ONE-TIME PASSWORD FROM MITM ATTACK**

   For doing online transaction, the customer has to provide user-id and pin number for authentication. A one-time password will be generated on the server machine using random number generator which is passed on to the customer's registered mobile no. In order to secure OTP from MITM attack the proposed solution is to generate a Transaction Password along with SMS OTP; rest of the methodology for authentication is same as mentioned above.

In this method Transaction Password is generated using OTP along with a secret key that has already been shared with the customer, which always remains the same. A mobile application is created and is installed on user's mobile device which does not require any Internet support.

The details that are preserved with server are shown in Table II as follows:

TABLE II: Login table to the main website with OTP and Transaction Password

| User name | One Time Password | Secret Key | Token | Status | Transaction Password |
|---|---|---|---|---|---|
| John | 891632 | 4321 | 1 | Valid | 17103118278 |
| Jim | 621589 | 4567 | 0 | Expired | 1881261730088 |
| Rusty | 123151 | 2234 | 1 | Valid | 1118271261 |
| David | 356123 | 3458 | 0 | Expired | 272680181827 |

When User receives OTP, he will enter this OTP and the secret key in the mobile application which will generate the transaction Password. Same process runs on the server to generate the transaction password. This transaction password is entered on the ecommerce website by the user for successful Transaction which is compared with server data for authentication.

**A. Steps for generating transaction password:**

**Step 1:** In order to generate the transaction password, two inputs are required by the user.
- 6 digits OTP generated using a random generator for example 891632.
- Secret key for example 4321.

**Step 2:** If we use the secret key as-is, it can be guessed by the hacker after a series of sniffed transactions. Hence, summation of secret key is used in our method. Summation is the operation of adding a sequence of numbers; the result is their sum or total. Here we take the summation of individual digits of the secret key. The summation for secret key given in step 1 is 10. This would make it more difficult for the hacker to decrypt the secret key.

**Step 3:** In order to generate the Transaction Password we combine OTP and summation of the secret key. However, if we keep the position of secret key summation static; it can be easily decrypted by the malicious users. Hence, we do not keep the position of summation of secret key constant. Thus, it is inserted each time at the location given by last digit of OTP.

For our example, the given OTP's last digit is 2, so location of secret key summation is 2. The remaining digits of the OTP are kept as continued in the same sequence after entering the secret key summation at the location specified by the last digit of OTP. Always OTP is 6 digits, if the last digit value is more than 6 suppose for example OTP is 621589 and summation of secret key is 22. Here 22 would be inserted at 9th location and middle positions are padded with zeros.

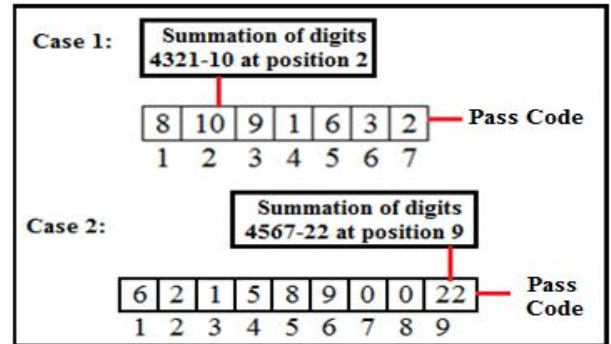

Fig 5: Example showing how OTP and summation of secret key combination is done to get the Pass Code

**Step 4:** Now we want to generate the transaction password through RSA algorithm. For this purpose, a constant 'bit strength' is used, which is the size of the RSA modulus in bits. The sub steps are given below [4]:

1. Choose two large prime numbers P and Q (We have assumed p=3 and q=11). Calculate N=P X Q (So the value of N=33) and Φ (n) = (p- 1)*(q-1) (So the value of Φ (n)=20)
   a. Compute the decryption key also known as private key defined as **d** such that gcd (d, Φ (n)) =1, where gcd () shows the greatest common divisor of two numbers.
   b. Now compute the encryption key also known as public key defined as **e** such that ed * mod (Φ (n)) =1, where mod is the remainder.
2. We have two keys, encryption and decryption keys e and d, respectively. Now, the pass code generated in step 3 is treated as the plain text defined as P and initially this pass code is encrypted using OTP and Secret key as shown in Figure 5. The plain text is converted into the cipher text represented as C and it is given by

$$C = P^e \bmod n$$

After performing the encryption, we get the value of C, which is the Transaction Password used for successful ecommerce Transaction. This is then used as an input for the decryption process.

3. The cipher text C is to be decrypted so that one could get again the Plain Text P which can authenticate server that Transaction password generated and used by the user is from the original pass code. The plain text i.e. original Pass code number which is a combination of OTP and Secure Pin is obtained by

$$P = C^d \bmod n$$

**B. Experimental Study of RSA with an Example:**

Let us start the experimental study by considering the Pass code 81091632 calculated as shown in the figure 5 after applying the above steps 1, 2 and 3. Now step 4 is used for generating the Transaction Password as computed below:

Assume the two small prime values for demonstration of the result i.e. p = 3, q = 11, therefore, n = p*q=3*11=33 and Φ (n) = (p- 1)*(q-1) = (3- 1)*(11-1) = 2*10 = 20. Now let us compute the encryption and decryption keys e and d, respectively. From step 1(a), gcd (d, 20) = 1 and select d=7 so that gcd (7, 20) = 1. Now, we have to choose the value of e such that e*d mod Φ (n) = 1 as per step 1(b). Therefore, 7*e mod (20) = 1 and it is valid when e=3. The complete encryption and decryption technique is shown in Table III in which original Pass code is used to generate transaction password using RSA Algorithm. The Pass code 81091632 is taken as a value of P for each digit and then after performing encryption Transaction Password generated is 17103118278, which is then used by the user for Ecommerce Transaction.

TABLE III: Encryption and Decryption of the Pass code Using RSA Algorithm to get the Transaction Password

| Digits of Pass Code | Encryption Module | | Decryption Module | |
|---|---|---|---|---|
| P | $P^3$ | $P^3$ mod 33=C | $C^7$ | $C^7$ mod 33=P |
| 8 | 512 | 17 | 410338673 | 8 |
| 1 | 1 | 1 | 1 | 1 |
| 0 | 0 | 0 | 0 | 0 |
| 9 | 729 | 3 | 2187 | 9 |
| 1 | 1 | 1 | 1 | 1 |
| 6 | 216 | 18 | 612220032 | 6 |
| 3 | 27 | 27 | 10460353203 | 3 |
| 2 | 8 | 8 | 2097152 | 2 |

## V. CONCLUSION

Today, we do many of our banking and shopping transactions by a simple click of mouse. While this means a lot of comfort, convenience and cost savings to us, it also exposes us to new-age risks like 'phishing'. The thieves are getting more and more sophisticated and learning new fraudulent techniques to acquire sensitive information such as usernames and passwords from bank or credit card customers. SMS-based One-Time Passwords (SMS OTP) were introduced to counter phishing and other attacks against Internet services such as online banking. Today, SMS OTPs are commonly used for authentication and authorization for many different applications. Recently, SMS OTPs have come under heavy attack. Through this research paper we are presenting a novel approach to combat the man-in-the-middle attacks for OTP. An approach is proposed where user will retrieve the one time password by SMS. The received OTP and secret PIN is fed to a mobile application to generate transaction password. This paper proposes a simple yet effective solution to counter man in the middle attack in order to secure your Ecommerce Transaction.